\documentclass[a4paper,11pt]{article}

\usepackage{jinstpub} 
\usepackage{hyperref}
\usepackage{lineno}
\natbibsorttrue  %% what does this do?

\newcommand{\DELTATSM}{1000} %%% SM coincidence must be at time > Tmu + DELTATSM, 

\newcommand{\TMLY}{99.5\pm 16.8} %%% best-fit LY and unc. for TM. Use in mathmode
\newcommand{\SMLY}{95.9\pm 11.7} %%%% best-fit LY and unc. for SM
\newcommand{\COMLY}{99\pm 15} %%%% combined LY for TM & SM

\title{\boldmath Performance of a Ton-scale  Water-based Liquid Scintillator Detector}

\author[a]{Rong Zhao,}
\author[b,1]{Lindsey Bignell,\note{Now at ARC Centre of Excellence for Dark Matter Particle Physics, Department of Nuclear Physics and Accelerator Applications, The Australian National University, Canberra, ACT 2601, Australia} }
\author[b]{David E. Jaffe,}
\author[b]{Richard Rosero,}
\author[b]{Minfang Yeh,}
\author[a]{Wei Wang,}
\author[b,2]{and Aiwu Zhang \note{Now at Leidos Inc., Vista, CA 92081}}

\affiliation[a]{Sun Yat-Sen University, Guangzhou, China}
\affiliation[b]{Brookhaven National Laboratory, Upton, New York, USA}

\emailAdd{zhaor25@mail2.sysu.edu.cn, djaffe@bnl.gov}
\abstract{
This study reports the performance and light yield of 1\% concentration water-based liquid scintillator (WbLS) deployed in a 1000-liter detector. 
A light yield of $\COMLY$ photons per MeV is determined by comparing data with simulation. 
This result aligns with our previous light yield determination using smaller detectors, thus establishing a solid foundation for the ongoing development and deployment of WbLS in larger-scale detectors. 
The feasibility of {\it in situ} preparation and the stability of light yield in WbLS are demonstrated, reinforcing its suitability for long-term experimental endeavors.

}


\begin{document}
\maketitle
\flushbottom

\section{Introduction}
\label{sec:intro}
The development and implementation of very large liquid-based particle detectors has led to significant advances in fundamental physics over the past decades~\cite{ref:SK,ref:SKD,ref:SNO,ref:KamLAND,ref:KamLANDN,ref:dyb,ref:dyba,ref:nova,ref:novals,ref:prospect,ref:prospectas,ref:prospectll}. 
Light production in these detectors is dominated by Cherenkov photons in water-based detectors or by scintillation photons in liquid-scintillator-based detectors. 
{In the 1980s, the concept of integrating the Cherenkov and scintillation processes in a water-based scintillator by introducing a fluor into water using a surfactant was attempted ~\cite{ref:wblsllc}. } 
{ More recently, the capability to produce water-based liquid scintillator (WbLS) by using a surfactant to emulsify organic liquid scintillator into water was demonstrated~\cite{ref:wbls_yeh}.} 
WbLS is attractive because the scintillation process for particles below the Cherenkov energy threshold enhances detector sensitivity. 
Due to the use of pure water rather than traditional oil-based solvents, it also provides a cost-effective and eco-friendly method for building large detectors.
 Recent studies have aimed to better understand the production and basic features of WbLS~\cite{ref:wbls_l,ref:wbls_ll,ref:wbls_j,ref:wbls_tr,ref:wbls_mp,ref:wbls_ply}. 
 The potential applications of WbLS have been discussed elsewhere~\cite{ref:wbls_asdc,ref:wbls_le,ref:wbls_css} with specific proposals of large scale ($\ge 10\ {\rm kiloton}$) detectors such as THEIA~\cite{ref:wbls_theia,ref:theia_s} and WATCHMAN~\cite{ref:wbls_watchman}.

At the Brookhaven National Laboratory (BNL), we built and operated a 1000-liter (``1-ton'') detector that was exposed to cosmic rays and viewed by eight 2-inch photomultiplier tubes (PMTs). 
The detector was initially filled with water before being converted to 1\%-by-mass WbLS with the addition of organic liquid and {\it in situ} mixing. 
A full GEANT4-based detector simulation~\cite{ref:geant} was conducted to calibrate the performance of the detector and determine the light yield (LY) of WbLS using a simulation model that defines the WbLS's light production, absorption, and re-emission~\cite{ref:wbls_l}. 
The long-term stability of WbLS was demonstrated with the months-long operation.  
Section~\ref{sec:exp_setup} describes the experimental setup and  WbLS fabrication. 
In Section~\ref{sec:sim_cali}, the simulation and calibration of the detector are presented. 
Section~\ref{sec:om_ly} describes the optical model of WbLS and the evaluation of the WbLS light yield.

%%%%%%%%%%%%%%%%%%%%%%%%%%%%%%%%%%%%%%%%%%%%%%%%%%%%%%%%%%%%%%%%%%%%%%%%%%%%%%%%%%%%%%%%%%%%%%%%%%%%%%
\section{Experiment setup}
\label{sec:exp_setup}
%   - detector geometry
\subsection{Detector geometry and components}
\begin{figure}
\centering
 \resizebox{0.365\textwidth}{!}{%
  \includegraphics{geoxy.pdf}%eps} 
} 
 \resizebox{0.365\textwidth}{!}{%
  \includegraphics{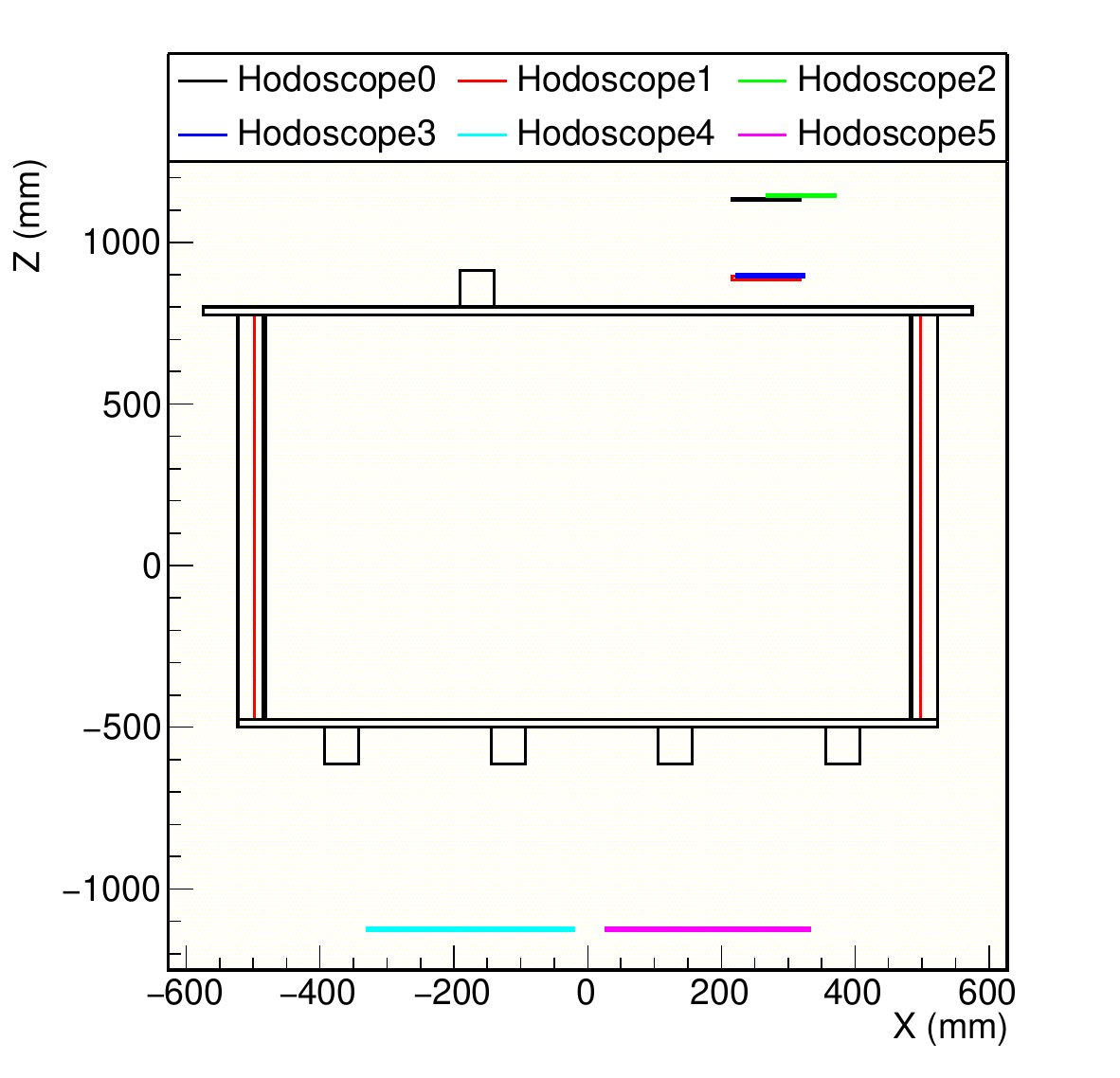}%eps} 
} 
   \resizebox{0.255\textwidth}{!}{%
\includegraphics{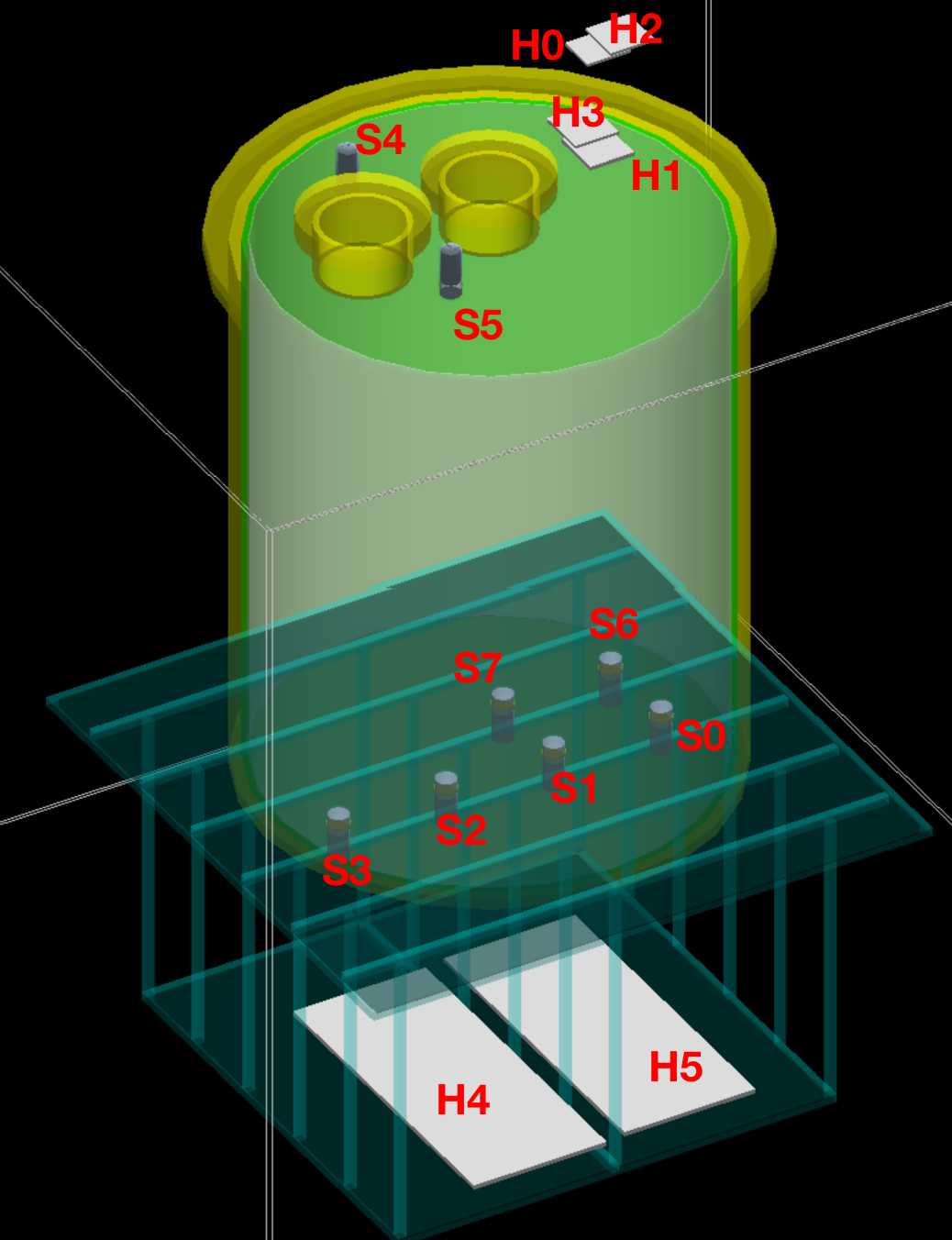} 
} 
\caption{Geometry of the 1-ton detector. Left panel: X-Y projection of the detector; middle panel:X-Z projection of the detector; right panel: detector visualization in GEANT4 simulation. 
The location of the signal PMTs (S0-S7) and hodoscopes are indicated.
{PMTs S4 and S5 are mounted on the upper lid; S0-S3 and S6-S7 are mounted on the base of the vessel.} }
\label{fig:det_geo}
\end{figure}

The 1-ton detector was a right cylindrical vessel made of 25.4 mm-thick, ultraviolet-transmitting  acrylic~\cite{ref:Nakano} with the cylinder axis aligned with the vertical $z$-axis as shown in Figure~\ref{fig:det_geo}.
The inner diameter and height of the main volume was 995 mm and 1250 mm, respectively, for a 972 liter capacity. 
As shown in Figure~\ref{fig:det_geo}, there was a central and off-axis port on the top lid, each with a 178 mm inner diameter and 3.1 liter volume. 
A 6.35 mm thick black PTFE sheet was installed inside the main volume covering the cylindrical surface to suppress reflections  on the vertical surface.  
The inner PTFE surface was roughened by brush sanding to minimize specular reflection.
{Multiple reflections from vessel surfaces complicated the interpretation of results in earlier studies~\cite{ref:wbls_l}; therefore, we opted to suppress  reflections from the vertical cylinder surface.}
The 1-ton detector was supported on an aluminum frame~\cite{ref:8020} in a dark room. 
Eight cylindrical 2-inch diameter PMTs~\cite{ref:PMTstory} on the top and bottom  viewed the internal volume of the vessel, labeled as S0, S1, S2...S7, and positioned as shown in Figure~\ref{fig:det_geo}. 
Optical cookies (2mm thick, EJ-560~\cite{ref:eljen}) were compressed between each PMT and the acrylic vessel to improve the transmission of optical photons produced in the liquid. 
 A system of four $\sim\!4$"$\times\sim\!4$" scintillator paddles defined a muon hodoscope above the top of the vessel, denoted as H0-H3. 
 Two large (12" $\times$ 28") scintillator paddles (H4 and H5) near the floor beneath the vessel were used to enhance the selection of samples of through-going muons (TM). 
{Ultrasonic} level sensors {(ToughSonic-3)} were installed in the two top ports to monitor the liquid level. 
 An LED with 410 nm central wavelength was located on the top of the lid between S4 and S5 for PMT calibration. 
 {The LED intensity was adjusted such that the average number of photoelectrons in each PMT was 0.1 or less.}
High purity nitrogen was flowed through the port volumes at $\sim\!0.5{\rm L}/{\rm min}$ to minimize contact between air and the liquid.

\subsection{Data acquisition and trigger system of the 1-ton detector}

The data acquisition (DAQ) system of the 1-ton detector was designed mainly to digitize PMT waveforms.
Signals from PMTs S0-S7 were passively split (0.89/0.11) with the larger signal sent to an FADC, CAEN V1729A, that was operated at 1 GSPS and acquired 2560 14-bit samples when triggered. 
For the ``HODO'' trigger, defined below, the cosmic ray muon signal appears around sample 130. 
The {smaller} signal is amplified and discriminated for triggering and online monitoring. 
Signals from H0-H5 are similarly amplified for monitoring and defining the ``hodoscope'' trigger. 
Two triggers were defined and used. 
 \begin{itemize}
     \item ``LED'': An externally {pulsed} LED at 0.5 Hz was used to provide single photoelectron (SPE) calibration for S0-S7.
     \item ``HODO'': This trigger was defined as $(\textrm{H2} + \textrm{H0}) \times (\textrm{H3}+\textrm{H1})$,  requiring a coincidence between at least one of the upper hodoscopes (H0, H2) and at least one of the lower hodoscopes (H1, H3). The HODO trigger rate was $\sim\!0.35\ {\rm Hz}$. 
 \end{itemize}
 
 \subsection{Liquid circulation system and preparation of WbLS}
 
A water circulation system was employed for filling and recirculation. 
For the initial fill, tap water was filtered and purified by reverse osmosis (RO) at about 10 liters per hour to fill the vessel over a 5-day period.
The recirculation system comprised deionization filtering and a degasser (SEPAREL EF-G5-B). 
The water was recirculated at 0.22 L/min using a peristaltic pump (ColePalmer 7528-10 with head 77200-62) to achieve a bubble-free liquid volume. 
Just before the acquisition of data used in the analysis reported in this paper, the peristaltic pump was replaced by a KNF Liquiport NF 300.TT 18S pump which enabled a higher circulation rate (up to 0.8 L/min) while presenting only fluoropolymer surfaces to the circulating liquid. 
Samples of the water exiting the 1-ton detector were periodically obtained. {The attenuation length of these samples was} measured in a UV-vis spectrophotometer (Shimadzu UV-1800) using a 10~cm long cell.
The water was measured to have a maximum attenuation length in excess of $\sim\!20$ m in the wavelength range 300-500 nm. 
During operation, the liquid level was monitored and topped-up with RO-filtered water to be at least 2.5 cm above the bottom of the top ports to avoid trapped gas bubbles below the lid of the 1-ton detector. 

WbLS with 1\% liquid scintillator concentration by mass was produced by {\it in situ} {sequential} mixing.
The feasibility and methodology of {\it in situ} {sequential}  mixing was established using a 60-liter prototype. 
 This methodology was then used to convert the 1-ton liquid from pure water to WbLS with a multistep procedure based on the prototype results.
\begin{enumerate}
    \item  
    A surfactant is mixed with liquid scintillator (linear alkylbenzene, LAB). 
     The surfactant enables the dissolution of liquid scintillator into water.
    \item  
    The mixture is gradually added to pure water and recirculated until it is thoroughly mixed. 
    \item 
    Anti-scattering material is injected into the recirculation stream. 
\end{enumerate}

The water recirculation system was modified for WbLS by replacing all tubing, fitting, and valves that contact {or can be in contact with} the WbLS with parts fabricated from fluoropolymers to enable {\it in situ} mixing. 
The deionization filtering and degasser were bypassed for WbLS recirculation. 
The calculated concentration of WbLS in kg/kg was $(1.18\pm0.02)\%$; slightly larger than the nominal 1\% because we neglected to take into account the volume of the black PTFE sheet.

\begin{figure}[!htbp]
\begin{center}
\resizebox{1.05\textwidth}{!}{%
\includegraphics{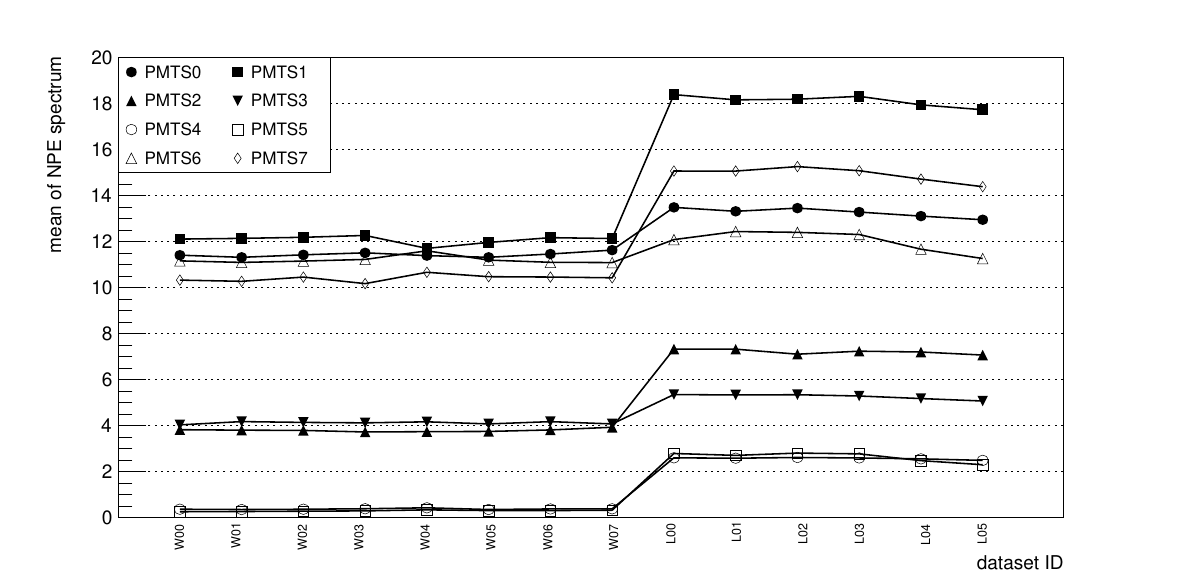}
}

\caption{Mean of the NPE spectrum for all signal PMTs with the {HODO$\times$H5} requirement. 
W00-W07 are the eight datasets with water and L00-L05 are the six datasets with WbLS.}
\label{fig:meannpe_period_all}
\end{center}
\end{figure}

Figure~\ref{fig:meannpe_period_all} shows the mean number of photoelectrons (NPE) for each signal PMT as a function of the dataset {for through-going muons selected by a coincidence between the HODO trigger and H5}.
{(PMT illumination for through-going and stopped muons is discussed in Section~\ref{sec:detsim}.)}

Datasets W00-W07 and L00-L05  were acquired 2 January - 25 June 2018 and 6 August 2018 - 28 January 2019, respectively. 
The {\it in situ} mixing occurred in July 2018.
Breaks between datasets correspond to interruptions of data acquisition, cycling of the PMT high voltage to access the interior of the dark room or changes in liquid circulation rates.
The increase in detected light for the WbLS data compared to the water is clear and the mean NPE for the water data is constant to within less than 10\%. 
For the latter WbLS datasets, a decline in mean NPE is observed.  

On 10 October 2018, about 1.2 L of water was added from an external water source, not via the RO-filter, to maintain the liquid level. 
After this addition, a monotonic increase in absorbance was observed for exit samples of the liquid. 
The reason for this increase was suspected to be an inadvertent introduction of impurities on 10 October.
We only use the WbLS datasets L00-L02 for analysis and exclude datasets L03-L05 due to the deteriorating performance. 
All water datasets W00-W07 are used for analysis.

 %%%%%%%%%%%%%%%%%%%%%%%%%%%%%%%%%%%%%%%%%%%%%%%%%%%%%%%%%%%%%%%%%%%%%%%%%%%%%%%%%%%%%%%%%%%%%%%%%%%%%%
\section{Detector simulation and calibration}
\label{sec:sim_cali}

\subsection{Waveform processing}
Acquired waveforms for S0-S7 were processed to select pulses. 
Denote the height and center of the $i^{\rm th}$ waveform sample as $h_i$ and $T_i$, respectively. 
Define the waveform baseline mean $m \equiv \sum_{i=A}^{B}h_i/(T_B-T_A)$ and variance $s^2 \equiv \sum_{i=A}^{B} (h_i - m)^2/(T_B-T_A-1)$, where $T_A=1560$ and $T_B=2560$. 
Contiguous samples with $h_i > m+5s$ were grouped and the largest sample of a group was selected to define the peak, $\Tilde{T}_i$. 
Pulses were defined by the sample range $(\Tilde{T}_i-5,\Tilde{T}_i+20)$. 
If two neighboring pulses overlapped, they were combined to form a single pulse. 
The total area of a pulse is referred to as the ``charge''. 

\subsection{Single photoelectron calibration}
Using LED trigger events, the SPE spectrum for PMT ${\rm S}_i$ was determined from $\sum_{i=a}^{b}h_i$ where $a = T_L - 5$ and $b = T_L +20$ with $T_L=$ the sample corresponding to the LED trigger. 
The spectrum was fitted {with a Poisson distribution convolved with a Gaussian} to determine the mean number of counts corresponding to a SPE which is called the PMT ``gain''. 
The PMT gain varied less than 2.6\% for the water and WbLS datasets. 
The measured SPE spectrum for each PMT was used in the simulation to emulate the electronics response to an incident photon. 

\subsection{PMT optical response calibration}\label{sec:oprescalib}

A number of wavelength-dependent factors influence the NPE detected by a PMT for a given sample of charged particles. 
These factors include the PMT quantum efficiency and collection efficiency, the transmission of light through the liquid, acrylic, optical cookie, PMT glass, and within the PMT,  the production of light in the liquid and non-liquid materials, the scattering and reflection of light, etc.
In our previous analysis~\cite{ref:wbls_l}, we compared the NPE distribution for a PMT observed on a sample of charged particles in water in data and simulation and adjusted the simulation so that it reproduced the NPE distribution observed in data.  
Here we employ a similar, but simplified, approach where we adjusted the simulation by a single factor for each PMT based on water data, then used the adjusted simulation to evaluate the light yield of the WbLS. 
We refer to this single factor determined for each PMT for each data sample as the ``calibration factor'' in the following.

Two different data samples were employed  to determine the PMT calibration factors and to evaluate the performance of WbLS based on the HODO trigger:
\begin{itemize}
    \item TM events are selected by requiring the HODO trigger in coincidence with either of the bottom hodoscopes H4 or H5. 
    \item Stopped muon (SM) events corresponding to events where an incident $\mu^\pm$ stops in the liquid volume and decays producing a $e^\pm$. 
    The selection criterion for SM events is 
    $$(\textrm{HODO})\times (NPE^{\textrm{total}}_{\mu}<40)\times (\textrm{DTC}).     $$
    where $NPE^{\textrm{total}}_{\mu}$ is the total photoelectrons (PE) observed in S0-S7 coincident with the HODO trigger and  DTC designates a delayed time coincidence between PMTs $S_i$ and $S_j$ with $i\ne j$ where $t_i > t_{\rm HODO} + \DELTATSM\ {\rm ns}$ and $|t_i - t_j|< 40\ {\rm ns}$. 
    For each signal PMT, $t_{\rm HODO}$ is defined as the mean arrival time of the pulse induced by a muon satisfying the HODO trigger.
    The $NPE^{\textrm{total}}_{\mu}<40$ requirement suppresses through-going muons. 
    The delayed time coincidence selects the decay $e^\pm$.  
    The delay of greater than {$\DELTATSM$ ns significantly suppresses afterpulses and signal reflections due to the small} impedance mismatch at the passive splitter.

\end{itemize}
We separately simulated TM and SM events for the water phase of the detector in order to calibrate the PMTs, followed by simulations for the WbLS phase of the detector in order to calculate the LY of WbLS.

\subsection{Detector simulation}\label{sec:detsim}
%%%%%%%%%%%%%%%%%%%%%
 A MC simulation using the RAT-pac framework~\cite{ref:rat-pac} was performed to better understand the energy deposition characteristics of cosmic rays as well as the production and propagation of photons in the detector.
 An optimal LY value was then determined by achieving the best agreement between the data and simulation NPE spectrum.
In this simulation, samples of cosmic ray muons {at sea level} were created using CRY~\cite{ref:CRY}.
The full momentum range of CRY-generated muons was used for the simulation of TM.
For SM, only muons with an initial momentum less than $500\ {\rm MeV}/c$ were generated to reduce CPU time.   
The generated muons were distributed over in a rectangle area ($1.0 \times 0.5\ {\rm  m}^2$) at a fixed height of about 100 mm above the top hodoscopes, H0 and H2. 
 In the simulation, the optical properties of all materials and optical boundaries (including the refractive index and attenuation) were taken from measurements or published data~\cite{ref:wbls_l}.
 In addition, the probability of diffuse reflection of the black PTFE was set to 3\% based on reflectivity measurements of similar black materials~\cite{ref:Marshall2014,ref:Schmidt2018}. 
We used a simple PMT optical model that assumed that all photons striking the photocathode could create photoelectrons. {A more complete optical model (GLG4) which took into account optical processes within the PMT was available in the simulation.
We used the GLG4 model to assess the uncertainty due to the PMT model.}
 Each simulated optical photon that was registered in the simulation was assigned a charge based on the measured SPE distribution for that PMT. 
 A total photoelectron count per PMT below 0.5 PE was assigned to be zero in both data and simulation to minimize the effects of the detection threshold, estimated to be 0.25 PE per PMT, in the data. 
 Only muons satisfying the HODO trigger logic were used for the simulation.

When the detector was filled with water, only Cherenkov photons produced by charged particles, mainly muons or decay electrons,  were detected. 
Due to the characteristic generation of Cherenkov light at fixed angle (about $41^\circ$ in water at 400 nm for $\beta\approx1$) from the charged particle trajectory, the PMTs on the bottom of the detector (S0-S3, S6, S7) were preferentially illuminated by TMs compared to PMTs S4 and S5 on the top as evident in Figure~\ref{fig:meannpe_period_all}.
Simulation shows that PMTs S4 and S5 were illuminated mainly by Cherenkov photons from large-angle delta rays produced by the TMs or by scattering of downward-going photons in the water. 
In addition, the simulation shows that the detected light for PMTs S1, S2 and S7 varied by up to 20\% if the TM traversed H4 or H5 due to Cherenkov photon generation in the acrylic. 
As evident from Figure~\ref{fig:meannpe_period_all}, the introduction of WbLS increased the overall detected NPE, with the most profound effects in the upper PMTs S4 and S5. 

In contrast to the downward illumination of TMs, decay electrons from SMs provided more uniform illumination of the PMTs albeit with reduced photon intensity due to the $\sim\!55\ {\rm MeV}/c$ electron momentum upper limit governed by muon decay.
Also in contrast to TMs, the decay electrons from SMs produced negligible light in acrylic according to simulation.

%%%%%%%%%%%%%%%%%%%%%%%%%%%%%%%%%%%%%%%%%%%%%%%%%%%%%%%%%%%%%%%%%%%%%%%%%%%%%%%%%%%%%%%%%%%%%%%%%%%%%%%%%%%%%
\subsection{Data analysis and calibration of the 1-ton detector}
\label{sec:data}

Approximately 880k (370k) valid HODO trigger events for water (WbLS) were recorded during the data-taking period of 2018-2019. 
Calibration factors for each PMT were determined using water data. 
Due to the differences in light production and detection, separate calibration factors were determined for the TM and SM samples. 

For TM, the total charge observed in each PMT in the time range $(t_{\rm HODO}, t_{\rm HODO} + 50\ {\rm ns})$ was {entered into histograms} for data and simulation.
For SM, the PMTs were divided into two sets dubbed ``EVEN'' (S0,2,4,6) and ``ODD'' (S1,3,5,7). 
If a DTC (Section~\ref{sec:oprescalib}) is found between two EVEN PMTs, then the total charge observed in each ODD PMT with time greater than $t_{\rm HODO}+\DELTATSM\ {\rm ns}$ was {entered into histograms}. 
Likewise, if a DTC between two ODD PMTs was found, the total charge in each ODD PMT was {entered into histograms}. 
Alternative selections, dubbed RED (S0,2,5,7)-BLUE (S1,3,4,6) and INNER (S1,2,5,7)-OUTER (S0,3,4,6) were defined and used for systematic studies. 
The total simulated NPE registered for the $j^{\rm th}$ PMT was scaled by the calibration factor $f_j$ prior to {entering into a histogram} and the area of each simulated distribution was normalized to the number of selected data events. 

For the $j^{\rm th}$ PMT, $\chi_j^2$ is formed
\begin{equation}\label{eqn:calib_chi2}
\chi_j^2 \equiv \sum_i^{\rm bins} \left( \frac{N_i^{\rm data} - N_i^{\rm MC}(f_j)}{\sigma_i}\right)^2
\end{equation}
\noindent where $N_i$ was the number of counts in the $i^{\rm th}$ bin and $\sigma_i$ was the total statistical uncertainty in the $i^{\rm th}$ bin. 
The total number of simulated events was about ten times that of data, so $\sigma_i$ was dominated by the data statistics.
In addition, to ensure a valid application of $\chi^2$, each data and simulation bin was required to have greater than 10 counts. 
This requirement was enforced by defining an ``overflow'' bin for each PMT  which contained the sum of all counts in that bin and above. 
We note that in both the TM and SM samples, more events in the overflow bin were observed in data than in simulation. 
For TM, we suspect this was due to simultaneous traversal of detector by multiple muons that was not adequately simulated.  

Using a simulation, the $\chi_j^2$ procedure was shown to produce an unbiased estimate of $f_j$ with two iterations if the true $f_j$ was in the range (0.2,2.2). 
The default simulation (that is, with $f_j=1 \forall j$) was run and best-fit $\tilde{f}_j$ were determined. 
A second simulation with $f_j = \tilde{f}_j$ was run and the unbiased, best-fit $\tilde{\tilde{f}}_j$ were determined. 

\subsection{Calibration results}
\label{sec:calires}

\begin{figure*}
 \resizebox{\textwidth}{!}{%
  \includegraphics{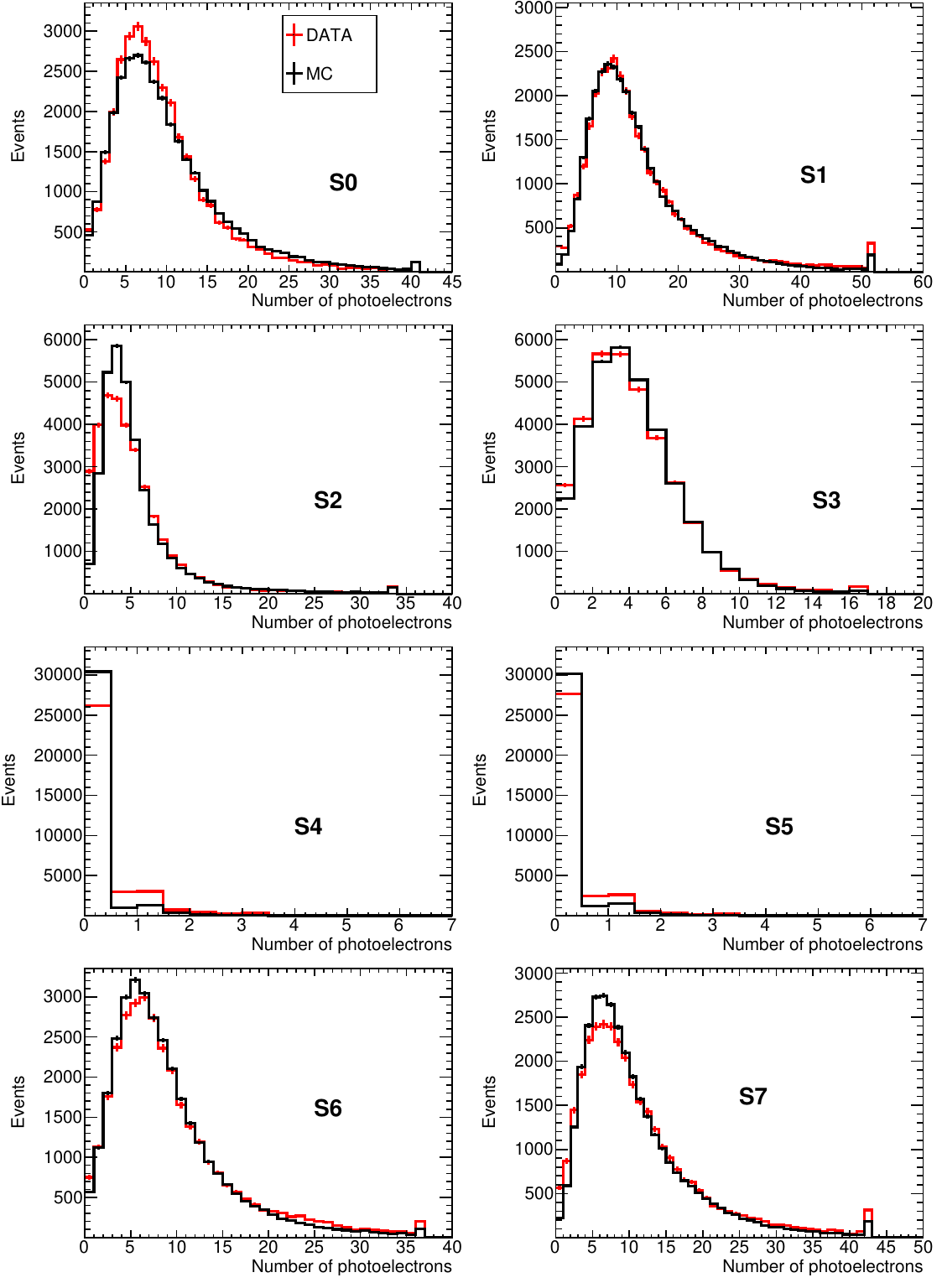}
 }
\caption{The NPE distribution for eight signal PMTs for the water data with calibration factors applied in the simulation, through-going muons. }
\label{fig:2npe}
\end{figure*}

\begin{figure*}
 \resizebox{\textwidth}{!}{%
  \includegraphics{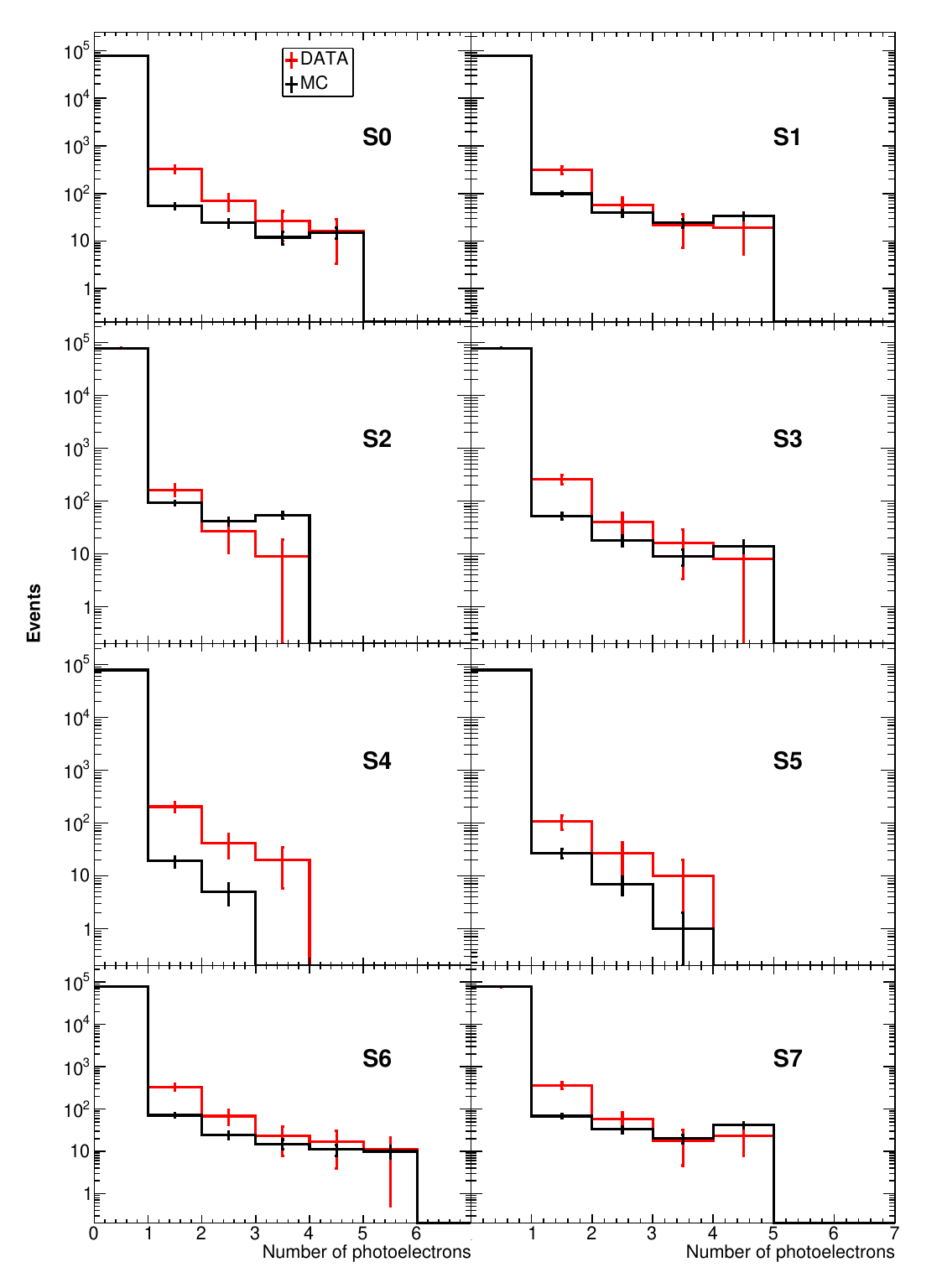}
 }
\caption{The NPE distribution for eight signal PMTs for the water data with calibration factors applied in the simulation, stopped muons. }
\label{fig:2npe_sm} 
\end{figure*}

The NPE distribution of PMTs with calibration factors applied for TM and SM events are shown in Figures~\ref{fig:2npe} and \ref{fig:2npe_sm}, respectively. 
About 33k (13k) water data events were used in the TM (SM) calibration. 
For TM calibration (Figure~\ref{fig:2npe}), a prescale factor of about 10 was used in the event selection. The agreement between the data and calibrated simulation is { acceptable} except for S4 and S5, the PMTs on the top, which have very low mean PE, and S2.
Agreement between data and calibrated simulation is acceptable for SM events, although the mean NPE is only about 0.5 for all PMTs.  

\begin{table}[!htbp]
  \caption[]{The calibration factors ($f$) and uncertainties ($\delta\!f$) as determined by TM and SM events.\label{tab:cf_error}}
   \centering
   \begin{tabular*}{.95\textwidth}{@{\extracolsep{\fill}}c|cccccccc}
   \hline\noalign{\smallskip}
   PMT &S0 & S1 &S2&S3&S4&S5&S6&S7\\
   \noalign{\smallskip}\hline\noalign{\smallskip}
   $f^{\textrm{TM}}$&$1.16$ & $0.73$ & $0.42$ & $0.95$ & $1.09$ & $1.07$ & $1.00$ & $0.81$ \\
   $\delta\!{f}^{\textrm{TM}}$&$0.09$ & $0.06$ & $0.03$ & $0.07$ & $0.31$ & $0.20$ & $0.06$ & $0.07$ \\ 
    \noalign{\smallskip}\hline\noalign{\smallskip}
   $f^{\textrm{SM}}$& $1.12$ & $0.90$ & $0.64$ & $1.05$ & $1.73$ & $1.65$ & $1.00$ & $0.88$ \\ 
   $\delta\!{f}^{\textrm{SM}}$&$0.08$ & $0.06$ & $0.04$ & $0.09$ & $0.34$ & $0.57$ & $0.07$ & $0.07$ \\
   \noalign{\smallskip}\hline
   \end{tabular*}
\end{table}

%%%%%%%%%%%%%%%%%%%%%%%%%%%%

The best-fit calibration factors of the eight PMTs for TM and SM are shown in Table~\ref{tab:cf_error}. 
The uncertainties include both statistical and systematic uncertainties.
Systematic uncertainties dominate. 
For the TM events, the systematic uncertainty of calibration factors was estimated by repeating the calibration procedure while individually varying simulation parameters, event selection criteria, fitting configurations, or datasets. 
In particular, the simulated horizontal ($xy$) positions of the hodoscopes were varied by their estimated uncertainties. 
The simple PMT optical model was replaced by the full GLG4 model. 
The estimated water, acrylic and optical cookie attenuation lengths in the simulation were varied. 
The event selection was varied by changing the definition of the HODO trigger, requiring different combinations of the 4 top plastic scintillator paddles. 
The fitting procedure was varied by omitting either the first or last (overflow) bin and by changing the bin width from 1.0 PE  to  0.5 PE. 
The dataset was changed from the full range of W00 through W07 to W00 and W07 only. 
The systematic uncertainty for the TM calibration was then taken as the RMS of the calibration factors for all the variations with respect to the nominal best fit. 

Analogous variations were used for the SM calibration, except that the event selection criteria was varied by using RED-BLUE or INNER-OUTER coincidences as described earlier or by removing the ${\rm NPE}^{\rm total}_{\mu}$ requirement. 
Similarly, the systematic uncertainty for SM calibration was taken as the RMS of the fitted calibration factors for all variations.  

The calibration factors for the top PMTs S4 and S5 in general have larger uncertainties since they are not as well illuminated. 
As mentioned earlier, the calibration factors for the two samples are not expected to be identical due to differences in light production and detection. 

%%%%%%%%%%%%%%%%%%%%%%%%%%%%%%%%%%%%%%%%%%%%%%%%%%%%%%%%%%%%%%%%%%%%%%%%%%%%%%%%%
\section{Optical model and light yield determination of WbLS}
\label{sec:om_ly}

A detailed WbLS model~\cite{ref:wbls_l} which took into account photon transmission processes (such as absorption and re-emission) was created based on experimental data obtained with ${\cal O}(1)$-liter liquid volumes. 
We used the same optical model as Ref.~\cite{ref:wbls_l} for the 1-ton simulation and considered the light yield as the only free parameter. 
The light yield of WbLS was determined by comparing the NPE spectra from data and simulation with the calibrated PMT response.

\begin{figure}[!htbp]
\begin{center}
 \includegraphics[width=.95\textwidth]{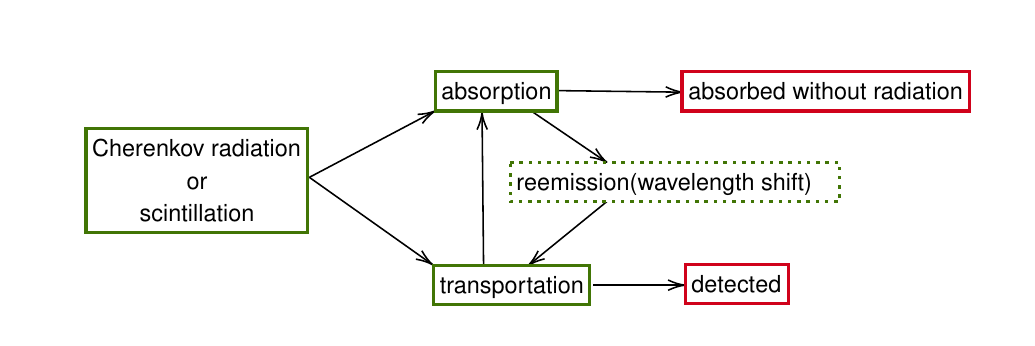}
\caption{Schematic of optical processes in WbLS optical model.
}
\label{fig:wbls_om}
\end{center}
\end{figure}
\begin{figure}[!htbp]
\begin{center}
 \includegraphics[width=\textwidth]{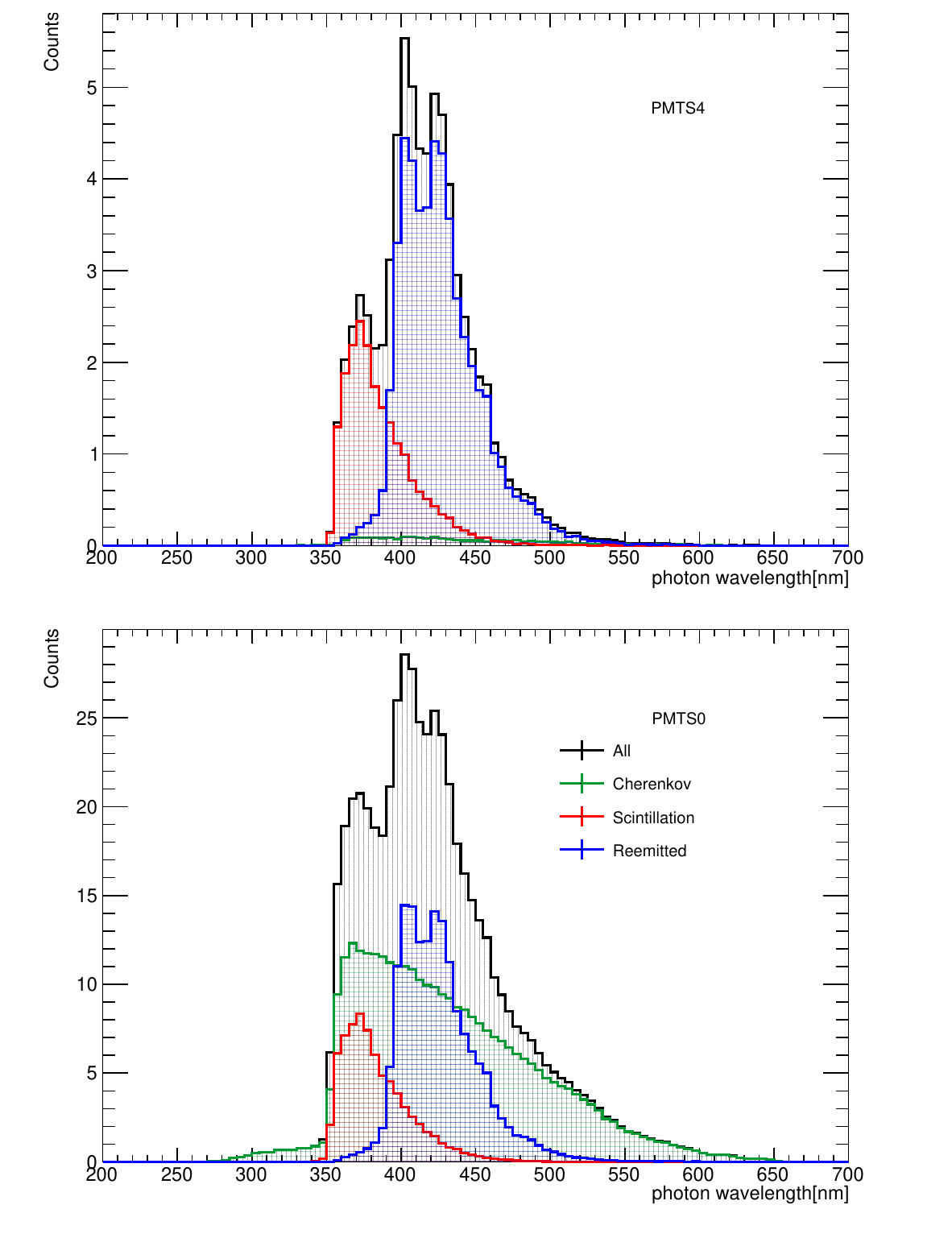}
\caption{The wavelength distribution of photons detected by S0 and S4 in the WbLS simulation with $LY=99$ optical photons per MeV for  TM events.}
\label{fig:wbls_wl}
\end{center}
\end{figure}

When the detector was filled with WbLS, photons generated by charged particles had a more complex transmission process than for pure water due to the absorption and re-emission mechanism of WbLS.  
These processes are shown schematically in Figure~\ref{fig:wbls_om}. 
Detectable photons from scintillation dominate over Cherenkov production in pure liquid scintillator (LS), while in 1\% WbLS, scintillation and Cherenkov light production are comparable due to the small proportion of LS. 
Absorption and re-emission of Cherenkov photons can significantly affect the total light yield since the re-emitted photons at longer wavelength may have a better match with the PMT wavelength response. 
The wavelength dependence of these effects in simulation with LY set to 99 optical photons per MeV is shown in Figure~\ref{fig:wbls_wl}.
%%%%%%%%%%%%%

The same data analysis algorithms and event selection criteria were used to determine the light yield of the WbLS optical model~\cite{ref:wbls_l} for WbLS data. 
For WbLS, a higher threshold for selecting stopped muon events is used, $\textrm{NPE}_{\mu}^{\text{total}}<60$, {which was determined by using the ratio of the mean NPE from WbLS and water data as the scaling factor.} 
For the TM and SM samples, the LY was determined by $\chi^2$-minimization of the difference between data and simulation for the NPE spectra of all eight PMTs taking into account the uncertainties in the per-PMT calibration factors.
Separate calibration factors (Table~\ref{tab:cf_error}) were employed for the TM and SM simulations as described previously. 
As with the determination of the calibration factors, an overflow bin is created to avoid any bin with less than 10 events. 
\begin{figure*}
 \resizebox{.86\textwidth}{!}{%
  \includegraphics{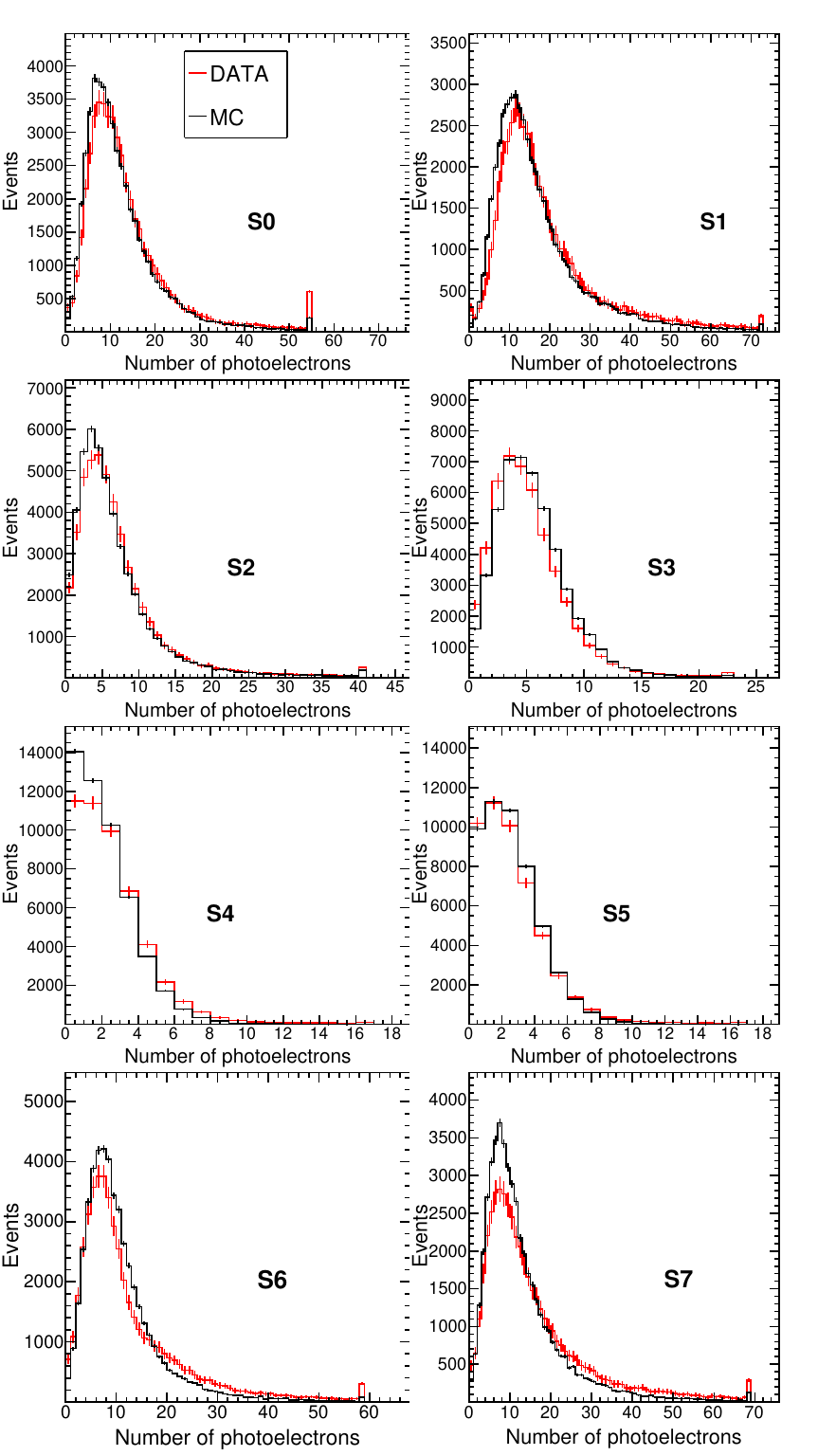}
 }
\caption{
The NPE distribution for eight signal PMTs for the WbLS data with optimal LY applied in the simulation for through-going muon events.}
\label{fig:wbls_npe_tm}  
\end{figure*}
\begin{figure*}
 \resizebox{\textwidth}{!}{%
  \includegraphics{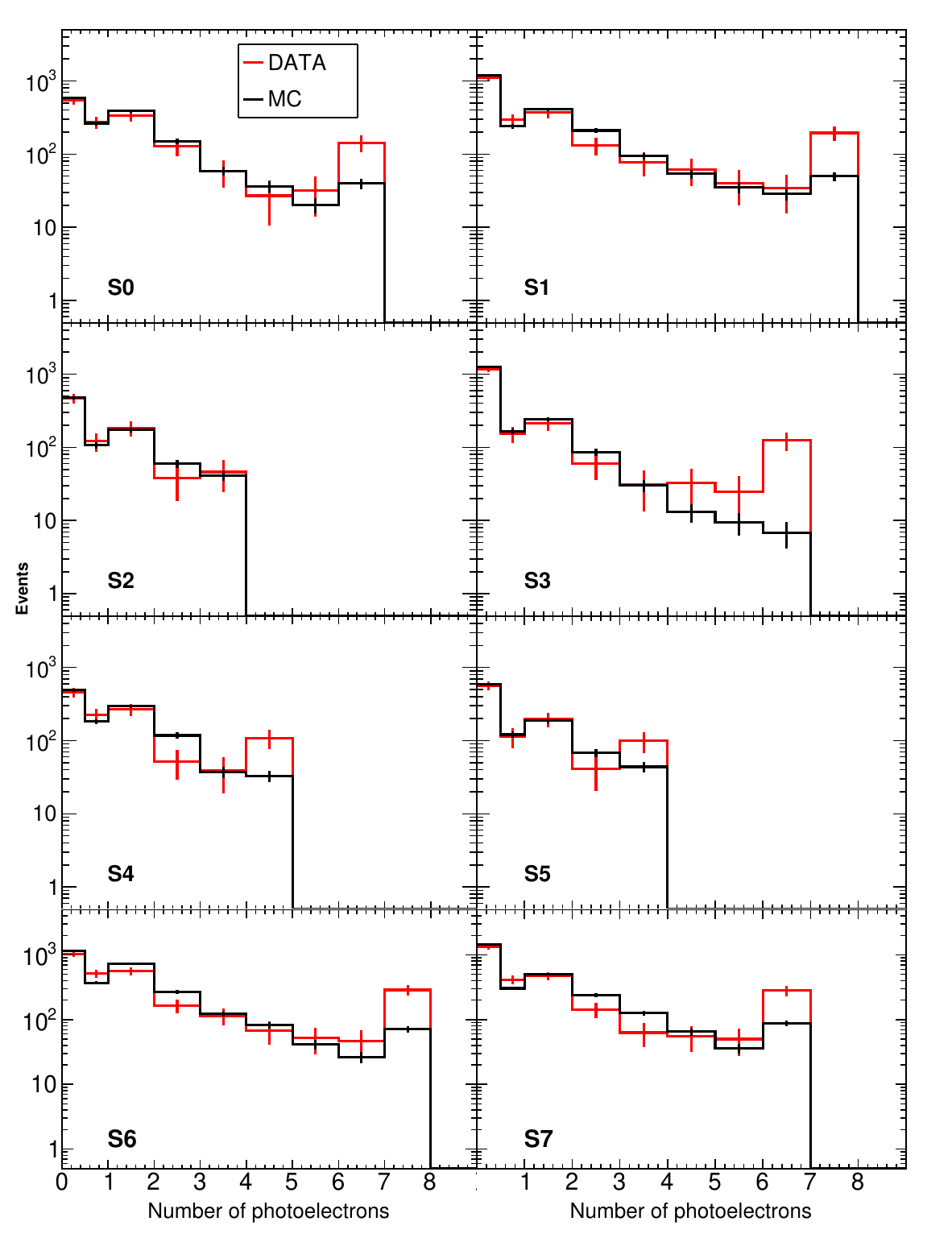}
 }
\caption{The NPE distribution for eight signal PMTs for the WbLS data with optimal LY applied in the simulation for stopped muon  events.}
\label{fig:wbls_npe_sm} 
\end{figure*}

The best-fit values of the LY are $\TMLY$ and $\SMLY$ optical photons per MeV for the TM and SM samples, respectively, where the uncertainty is the combined statistical and systematic uncertainty. 
The systematic uncertainty takes into account a variation in the WbLS attenuation length by $\times \frac{1}{2}$ and $\times 2$ in the optical model, the contribution of $\chi^2$ from top PMTs, S2 and first bins, overflow bins in the NPE spectrum, bottom hodoscope and time coincidence group and total NPE cut threshold.
The LY determined with the TM and SM samples was $\COMLY$ optical photons per MeV in agreement with $108.9\pm10.9$ that was determined with ${\cal O}(1)$-liter volumes in our previous work~\cite{ref:wbls_l}. 
Figure~\ref{fig:wbls_npe_tm} (Figure~\ref{fig:wbls_npe_sm}) displays the TM (SM) NPE {spectra  of the signal PMTs for WbLS data compared to simulation with the best-fit LY.} 
%%%%%%%%%%%%%%%%%%%%%%%%%%%%%%%%%%%%%%%%%%%%%%%%%%%%%%%%%%%%%%%%%%%%%%%%%%%%%%%%%%%%%%%%%%%%%%%%%%%%%%
\section{Conclusion}

In this study, we designed and operated a $\sim\!\!1000$ liter vessel filled with water-based liquid scintillator.
We confirmed the feasibility of {\it in situ} {sequential} WbLS preparation and demonstrated its months-long stability of WbLS at a concentration of $\sim\!\!1\%$ by mass. 
Based upon a detailed WbLS model~\cite{ref:wbls_l} in a comprehensive detector simulation, we measured consistent light yield per energy deposited for samples of through-going and stopped muons of $\TMLY$ and $\SMLY$ optical photons per MeV, respectively. 
Combining these results, we determined the light yield of WbLS to be $\COMLY$ optical photons per MeV, consistent with a previous light yield determination using \mbox{${\cal O}(1)$-L} detectors. 
These results provide confidence in the ability to produce and operate large scale WbLS-based detectors with properties consistent with small prototypes.

\acknowledgments

This research was funded by Brookhaven National Laboratory LDRD 12-033 and by U.S. Department of Energy Grant No. DE-SC0012704. 
We thank Harry Themann for the technical design of the 1-ton mechanical support and dark enclosure. 
We acknowledge and thank Russell Burns for technical assistance and contributions to the recirculation, gas and mechanical systems.

\end{document}